\documentclass[final,3p,times]{elsarticle}
\usepackage{graphicx}
\usepackage{bm}
\newcommand{\bea}{\begin{eqnarray}}
\newcommand{\eea}{\end{eqnarray}}

\journal{Elsevier}

\begin{document}

\begin{frontmatter}

\title{Covariant propagator and chiral power counting}

\author{Zhou Liu}\author{Li-Hong Wen}\author{Ji-Feng Yang\corref{cor1}}

\address{School of Physics and Electronic Science, East China Normal University, Shanghai 200241, China}
\cortext[cor1]{Corresponding author}

\begin{abstract}Some one-loop diagrams with one and two external baryons/nucleons are revisited using covariant baryon propagators in chiral effective theory. We showed that it is enough to separate and subtract all the local terms that violate chiral power counting to recover chiral power counting, no need to introduce extra operations. The structures of leading chiral corrections and IR enhancement or threshold effects are 'stable' or persist as long as covariant propagators are employed for all particles.
\end{abstract}

\begin{keyword}
chiral effective theories \sep covariant propagator \sep power counting
\end{keyword}
\end{frontmatter}


\section{Introduction}\label{intro}
Chiral perturbation theory is a beautiful effective theory of QCD at low energies that works perfectly in pure pion sector. In the one baryon sector, it is convenient to adopt the heavy  baryon (HB) formalism\cite{Isgur,Jenkins} for many topics\cite{Bernard}. Since Weinberg's seminal paper in 1990\cite{Weinberg} and the pioneering works\cite{vK}, chiral perturbation theory has been extended to nuclear systems with significant successes\cite{RevB,RevBvK,RevEpel,RevRMP81,RevRP503}. Nevertheless, things become rather complicated when baryons are involved in loops\cite{GSS}: If one works with covariant baryon propagators, baryon masses contribute extra large scales that upset the standard chiral power counting\cite{GL,GL2}. In spite that it is enough to employ the HB formalism in one baryon sector to preserve chiral power counting, there do be occasions where it is flawed (mainly due to the loss of correct kinematics controlled by relativistic propagators) and not trustworthy\cite{becher}. This complication persists in the chiral effective theories for nucleonic systems or nuclear forces and has been intensively studied, see, e.g., Refs.\cite{PVRA,HPVRA,EpelMeis,EG,MLEA,LongYang,ZME}. Of course, the nonperturbative nature of nuclear system due to the large scattering lengths, etc., renders things even more difficult.

In recent years, there appear some evidences in favor of using covariant formulation of baryonic sectors\cite{EG716,MPLA29} where the key observation is that the large items that violate the standard chiral power counting are local and hence could be subtracted away by adopting an extended on-mass-shell (EOMS) prescription\cite{Gege-Scherer}. The recent efforts in constructing nuclear forces from a covariant approach show encouraging prospects\cite{EG716,Beihang}. In our view, one could work in a very convenient and unified manner for all the cases involving heavy hadron propagators in chiral loops: merely isolating the {\em local large (non-chiral) terms} and remove them via local counter terms. The retained chiral terms are what one would have in HB formulation. For multi-baryonic processes, there may be nonlocal large terms due to 'infrared enhancement' and/or threshold effects that become singular in HB formalism\cite{becher} and incur extra power like divergences due to non-relativistic treatment, which could only be definite within covariant formulation\cite{MPLA29}.

In this report, we wish to revisit some one-loop diagrams involving one and two baryons/nucleons to examine the chiral structures in a unified perspective. These diagrams will be recalculated using covariant baryonic propagators throughout the whole work, first with covariant vertices and then with HB vertices and composite operators. The standard chiral power counting is indeed revived after removing some local pieces. Moreover, provided that covariant propagators are employed, different version of pion-baryon interactions do not affect the leading chiral contributions (and the possible IR enhanced terms) after the large local pieces are separated and removed. This 'stability' of the chiral components suggests a extended subtraction scheme: First separate and remove all the local large terms due to baryon/nucleon masses, then subtract the remaining chiral divergences. Generically, one may anticipate that in covariant formulation for heavy hadrons (including bosons), all the local non-chiral (large) terms warrant an extended subtraction according to the spirit of decoupling\cite{dec,dec2,dec3}, a prescription that naturally preserves chiral power counting, while dispensing extra manipulations and automatically covering the sector with purely chiral degrees. In this sense, we arrived at a unified scheme for chiral effective theories for hadronic and nuclear physics. The nonperturbative treatment or resummation of certain definite items, according to our view, should also be pursued in covariant or relativistic formulation.

This report is organized as follows: In Sec. 2 we present the set up for our investigation. In Sec. 3 we present our main analysis of the one-loop diagrams calculated in covariant and mixed formulation, demonstrating the detailed chiral properties of these amplitudes. The summary is given in Sec. 4.
\section{Theoretical setup}Here we present the main tools to be used and the objects to be examined. We will consider diagrams with two and four external lines of nucleons for simplicity. As a low-energy effective theory of QCD, chiral perturbation theory has been first given in covariant formulation in history\cite{GSS}. In this work, we focus on the SU(2) chiral effective field theory. The covariant Lagrangian we work with reads (see Ref.\cite{RevRP503})\bea\mathcal{L}_{\chi ET}&=&\mathcal{L}_{\pi\pi}+\mathcal{L}_{\pi{N}}+\cdots,\\ \mathcal{L}_{\pi\pi}&=&\frac{1}{2}\partial_\mu\bm{\pi}\cdot\partial^\mu\bm{\pi}-\frac{1}{2}m^2_\pi\bm{\pi}^2+\mathcal{O}\left(\bm{\pi}^4\right),\\\mathcal{L}_{\pi{N}}&=&\bar{\Psi}_N \left[i\!\!\not\!\partial-M_N+\frac{g_{A}}{2f_\pi}\gamma^5\bm{\tau}\cdot\!\!\not\!\partial\bm{\pi}-\frac{1}{4f^2_\pi}\bm{\tau}\cdot\left(\bm{\pi}\times\!\!\not\!\partial\bm{\pi}\right) +\mathcal{O}\left(\bm{\pi}^3\right)\right]\Psi_N.\eea The HB form of $\pi-N$ interaction reads\bea{\mathcal{L}}_{HB,int}=\bar{\Psi}_v\left[-\frac{g_{A}}{2f_\pi}S^\mu\bm{\tau}\cdot \partial_\mu\bm{\pi}-\frac{1}{4f^2_\pi}\bm{\tau}\cdot\left(\bm{\pi}\times v^\mu\partial_\mu\bm{\pi}\right)+\mathcal{O}\left(\bm{\pi}^3\right)\right]{\Psi}_v,\eea where $S^\mu\equiv\frac{i}{2}\gamma^5\sigma^{\mu\nu}v_\nu=-\frac{1}{2}\gamma^5\left(\gamma^\mu\!\!\not\!v-v^\mu\right),v^2=1,\Psi_v(x)=e^{iM_Nv\cdot x}P^+_v\Psi_N(x), P_v^\pm\equiv\frac{1}{2}\left(1\pm\not\!v\right).$

In Refs.\cite{JWCJi,Arndt}, it is shown that in low momentum transfer the chiral properties or non-analytical quark mass dependence of the matrix elements of traceless twist-2 operators ($O^{a,(n)}_{\mu_1\mu_2\cdots\mu_{n}}(quark)=\frac{1}{n!}\bar{q}\tau^a\gamma_{\{\mu_1}iD_{\mu_2}\cdots{iD}_{\mu_n\}}q-traces$, $D$ applied on both $q$ and $\bar{q}$) between hadronic states could be computed through chiral perturbation theory in HB formulation through matching on to hadronic operators: $O^{a,(n)}_{\mu_1\cdots\mu_{n}}(quark)\rightarrow{O}_{\mu_1 \cdots\mu_{n}}^{a,(n)}(\pi)+{O}_{\mu_1\cdots\mu_{n}}^{a,(n)}(N),\ O_{\mu_1\cdots\mu_{n}}^{a,(n)}(\pi)=a^{(n)}2i^n\varepsilon^{abc}\pi^b\partial_{\mu_1}\cdots\partial_{\mu_n}\pi^c +\mathcal{O}\left({\bm{\pi}^3}\right)-traces,\ O_{\mu_1\cdots\mu_{n};HB}^{a,(n)}(N)=A^{(n)}\bar{\Psi}_v\left\{v_{\mu}\cdots{v}_{\mu_n}\frac{1}{2}\left(\xi\tau^a\xi^\dagger+\xi^\dagger \tau^a\xi\right)\right\}\Psi_v+\mathcal{O}\left(\bm{\pi}^3\right)-traces$, where $\xi\equiv\exp(i\bm{\tau}\cdot\bm{\pi}/2f_\pi)$, $a^{(n)}$ and $A^{(n)}$ being matching coefficients. In this work we will illustrate our points with the simple case $n=1$, namely, the diagrams with following composite operators:\bea&&O_{\mu}^{a,(1)}(\pi)=2a^{(1)}i\varepsilon^{abc}\pi^b \partial_{\mu}\pi^c+\mathcal{O}\left({\bm{\pi}^3}\right),\\&&{O}_{\mu;Cov}^{a,(1)}(N)=A^{(1)}\bar{\Psi}_N\left\{\gamma_{\mu}\left[\tau^a\left(1-\frac{1}{2f^2_\pi}\bm{\pi}^2\right) +\frac{1}{2f^2_\pi}\pi^a\bm{\tau}\cdot\bm{\pi}+\frac{g_A}{f_\pi}\gamma^5\left(\bm{\tau}\times\bm{\pi}\right)^a\right]\right\}\Psi_N+\mathcal{O}\left(\bm{\pi}^3\right),\\&&O_{\mu;HB} ^{a,(1)}(N)=A^{(1)}\bar{\Psi}_v\left\{v_{\mu}\left[\tau^a\left(1-\frac{1}{2f^2_\pi}\bm{\pi}^2\right)+\frac{1}{2f^2_\pi}\pi^a\bm{\tau}\cdot\bm{\pi}\right]+\frac{2g_A}{f_\pi}S_\mu \left(\bm{\tau}\times\bm{\pi}\right)^a\right\}\Psi_v+\mathcal{O}\left(\bm{\pi}^3\right).\eea Here, we note that the third term with single pion is not listed out in Ref.\cite{Arndt} as it does not contribute in the HB formulation. However, it does contribute in covariant formulation, which is necessary to match with the HB results, as will be shown below.

The specific diagrams to be recalculated are listed in Fig. 1, while the last two diagrams (i.e., (f) and (g)) actually do not contribute in the HB formulation. The matrix elements of such operators ($n=1$) between nucleon states will be denoted as\bea\mathcal{M}^{a,(1)}_\mu\left(p,p\right)\equiv\langle N(p)|{O}_{\mu;\cdots}^{a,(1)}(N)|N(p)\rangle,\quad p^2=M_N^2.\eea
\begin{figure}[t]\begin{center}\label{Fig1}\resizebox{15cm}{!}{\includegraphics{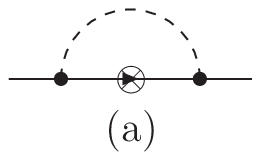}\includegraphics{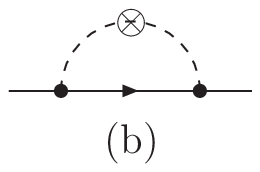}\includegraphics{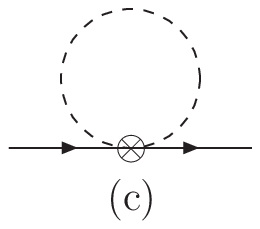}\includegraphics{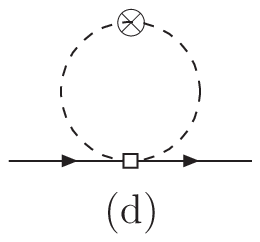} \includegraphics{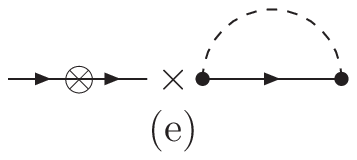}\includegraphics{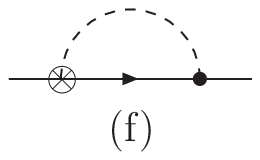}\includegraphics{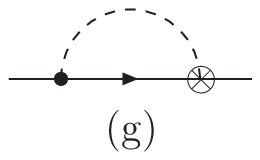}}\caption{One-loop diagrams for chiral corrections to matrix elements of the twist-2
operators (Eqs.(5,6)) in nucleon states}\end{center}\end{figure}

We will also recalculate the one-loop diagrams for $NN$ scattering listed in Fig. 2, namely the triangle and box diagrams for our purpose, the foot-ball diagram will not be considered as it does not contain baryon lines. To focus on our main points, we put that $p=\tilde{p}=p'=\tilde{p}'=(M_N,\mathbf{0})$:\bea\mathcal{M}_{NN}\left(p,\tilde{p};p',\tilde{p}'\right) \equiv\langle N(p),N(\tilde{p})|\mathcal{T}_{\chi ET}|N(p'),N'(\tilde{p}')\rangle.\eea
\begin{figure}[t]\begin{center}\label{Fig2}\resizebox{11cm}{!}{\includegraphics{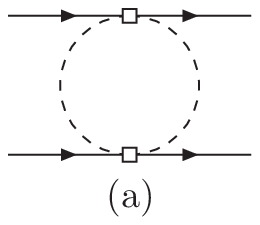}\includegraphics{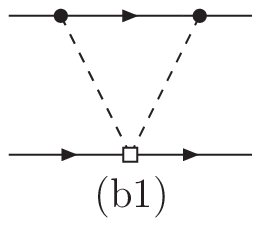}\includegraphics{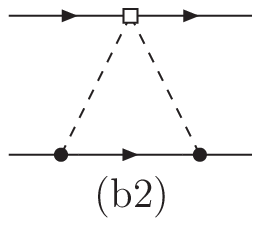}
\includegraphics{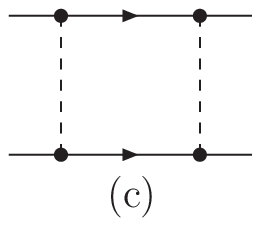}\includegraphics{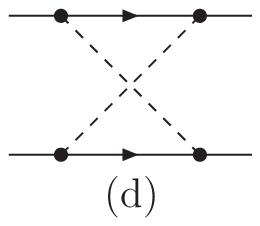}}\caption{One-loop diagrams for $NN$ scattering}\end{center}\end{figure}
\section{Calculation and Analysis}
\subsection{Covariant and mixed calculations}The detailed results of completely covariant calculation are listed in the Appendix A, where the following shorthand notations have been employed:\bea\label{def}\rho\equiv\frac{M^2_N}{m^2_\pi},\ \sigma\equiv\sqrt{4\rho-1},\ \ell_N\equiv\ln\frac{4\pi\mu^2}{M^2_N},\ \ell_\pi\equiv\ln\frac{4\pi\mu^2}{m^2_\pi},\ \Gamma_\epsilon\equiv\Gamma(\epsilon).\eea Here, we make some remarks about the results: First, it is straightforward to see that by requiring that $A^{(1)}=a^{(1)}=1$, the items from Fig.1 constitute the one-loop corrections to the vector current operator sandwiched between nucleon state, hence sum up to zero, i.e., the current is not renormalized as is the usual case for a conserved current:\bea\sum_{i=a}^{g}\mathcal{M}^{a,(1)}_{\mu;(i)}=0.\eea Like Fig. 2(a), Fig. 1(c) and Fig. 1(d) are completely chiral and will not be considered from now on. Second, there is a nonlocal and hence definite 'large piece' $\sim-\frac{12\rho}{\sigma}\arctan\sigma$ in Fig. 2(a) due to IR enhancement, which is also the dominant contribution to the contact $NN$ interaction $C_0(\bar{N}N)^2$\cite{MPLA29}. This term remains definite in the mixed calculation, see Appendix B.

In the mixed calculation using covariant propagators and HB vertices and composite operators (Appendix B), vector current conservation is no longer valid\bea\sum_{i=a}^{g}\mathcal{M} ^{a,(1);\texttt{\tiny HBv}}_{\mu;(i)}=-\bar{u}\tau^av_\mu{u}\frac{5g^2_Am^2_\pi}{16(4\pi f_\pi)^2}\left\{\frac{3}{2}\rho\left(\Gamma_\epsilon+\ell_N+\frac{13}{30}\right) +\left(\Gamma_\epsilon+\ell_N+\frac{3}{5}\right)+\frac{4}{5}\left(\mathcal{R}_{1a}^{\texttt{\tiny HBv}}-\mathcal{R}_{1b}^{\texttt{\tiny HBv}}\right)\right\}\neq0.\eea Obviously, the main sources of violation in the mixed calculation are the local large or non-chiral pieces. In other words, the chiral terms still preserve vector current conservation! This observation suggests that we should subtract all the local non-chiral terms, then canonical relations like current conservation would be preserved in chiral limit ($\rho\rightarrow\infty$):\bea\label{mix-subtract}\sum_{i=a}^{g}\mathcal{M}^{a,(1);\texttt{\tiny HBv}}_{\mu;(i);subtracted}=-\bar{u}\tau^av_\mu{u}\frac{5g^2_Am^2_\pi}{16(4\pi f_\pi)^2} \left\{\frac{4}{5}\left(\mathcal{R}_{1a}^{\texttt{\tiny HBv}}-\mathcal{R}_{1b}^{\texttt{\tiny HBv}}\right)\right\}=o(\rho^{-1/2}).\eea

For a more transparent analysis, it is instructive to decompose the amplitudes into terms of different chiral status, as will be done in next subsection.
\subsection{Decomposition and separation}To explore the properties of these one-loop diagrams, we decompose each of them into four categories: local items that are non-chiral, local item that is chiral, nonlocal relativistic corrections (i.e., the $\mathcal{R}$ terms) that vanish as $o(\rho^{-1/2}$) in the limit $\rho\rightarrow\infty$, and possible nonlocal 'large' items that arise from IR enhancement\cite{Weinberg} or threshold effects\cite{becher} and could not be subtracted by local counterterms, hence termed 'anomalous' here. To this end, we need the following crucial identity that is a simple consequence of Eq.(\ref{def})\bea\ln\rho=\Gamma_\epsilon+\ell_\pi-\left(\Gamma_\epsilon+\ell_N\right)\eea for terms tied with powers $\rho^{k},\ k=0,1$.

The results from covariant calculation (Appendix A) are then decomposed as below:\bea\mathcal{M}^{a,(1)}_{\mu;1(a)}&=&-A^{(1)}\bar{u}\tau^a\gamma_\mu{u}\frac{g^2_Am^2_\pi}{4(4\pi f_\pi)^2}\left\{\underline{2\left(\Gamma_\epsilon+\ell_N+\frac{3}{2}\right)}\overbrace{-3\left(\Gamma_\epsilon+\ell_\pi\right)}+\mathcal{R}_{1a}^{\texttt{\tiny c}}\right\}=\frac{1}{3} \mathcal{M}^{a,(1)}_{\mu;1(e)},\\\mathcal{M}^{a,(1)}_{\mu;1(b)}&=&a^{(1)}\bar{u}\tau^a\gamma_\mu{u}\frac{g^2_Am^2_\pi}{(4\pi f_\pi)^2}\left\{\underline{2\rho\left(\Gamma_\epsilon +\ell_N+1\right)+4\left(\Gamma_\epsilon+\ell_N+\frac{7}{4}\right)}\overbrace{-3\left(\Gamma_\epsilon+\ell_\pi\right)}+\mathcal{R}_{1b}^{\texttt{\tiny c}}\right\},\\ \mathcal{M}^{a,(1)}_{\mu;1(f)}&=&-A^{(1)}\bar{u}\tau^a\gamma_\mu{u}\frac{g^2_Am^2_\pi}{(4\pi f_\pi)^2}\left\{\underline{\rho\left(\Gamma_\epsilon+\ell_N+1\right)+\left(\Gamma_\epsilon +\ell_N+2\right)}+\mathcal{R}_{1f}^{\texttt{\tiny c}}\right\}=\mathcal{M}^{a,(1)}_{\mu;1(g)},\\\mathcal{M}_{NN;2(b1)}&=&\langle\mathbf{\tau}_1\cdot\mathbf{\tau}_2\rangle \frac{g^2_Am^2_\pi}{128\pi^2f^4_\pi}\left\{\underline{2\rho\left(\Gamma_\epsilon+\ell_N+1\right)+4\left(\Gamma_\epsilon+\ell_N+\frac{7}{4}\right)}\overbrace{-3\left(\Gamma_\epsilon +\ell_\pi\right)}+\mathcal{R}_{2b}^{\texttt{\tiny c}}\right\}=\mathcal{M}_{NN;2(b2)},\\\mathcal{M}_{NN;2(c)}&=&-\left(3-2\langle\mathbf{\tau}_1\cdot\mathbf{\tau}_2\rangle\right) \frac{g^4_Am^2_\pi}{128\pi^2f^4_\pi}\left\{\underline{\rho\left(\Gamma_\epsilon+\ell_N+3\right)-4\left(\Gamma_\epsilon+\ell_N+\frac{13}{16}\right)}\overbrace{+\frac{15}{4} \left(\Gamma_\epsilon+\ell_\pi\right)}\right.\nonumber\\&&\left.\underbrace{-\frac{12\rho}{\sigma}\arctan\sigma}+\mathcal{R}_{2c}^{\texttt{\tiny c}}\right\},\\\mathcal{M}_{NN;2(d)} &=&-\left(3+2\langle\mathbf{\tau}_1\cdot\mathbf{\tau}_2\rangle\right)\frac{g^4_Am^2_\pi}{128\pi^2f^4_\pi}\left\{\underline{3\rho\left(\Gamma_\epsilon+\ell_N+\frac{1}{3}\right) +4\left(\Gamma_\epsilon+\ell_N +\frac{17}{16}\right)}\overbrace{-\frac{15}{4}\left(\Gamma_\epsilon+\ell_\pi\right)}+\mathcal{R}_{2d}^{\texttt{\tiny c}}\right\}.\eea

Similarly, for those from mixed calculation (Appendix B) we have\bea\mathcal{M}^{a,(1);{\texttt{\tiny HBv}}}_{\mu;1(a)}&=&-A^{(1)}\bar{u}\tau^av_\mu{u}\frac{g^2_Am^2_\pi}{4(4\pi f_\pi)^2}\left\{\underline{\frac{5}{2}\rho\left(\Gamma_\epsilon+\ell_N+\frac{3}{5}\right)+\frac{15}{4}\left(\Gamma_\epsilon+\ell_N+\frac{5}{3}\right)}\overbrace{-3\left(\Gamma_\epsilon +\ell_\pi\right)}+\mathcal{R}^{\texttt{\tiny HBv}}_{1a}\right\},\\\mathcal{M}^{a,(1);{\texttt{\tiny HBv}}}_{\mu;1(b)}&=&a^{(1)}\bar{u}\tau^a v_\mu{u}\frac{g^2_Am^2_\pi}{(4\pi f_\pi)^2} \left\{\underline{\frac{5}{8}\rho\left(\Gamma_\epsilon+\ell_N+\frac{11}{10}\right)+\frac{5}{2}\left(\Gamma_\epsilon+\ell_N+\frac{11}{5}\right)}\overbrace{-3\left(\Gamma_\epsilon +\ell_\pi\right)}+\mathcal{R}^{\texttt{\tiny HBv}}_{1b}\right\}\nonumber\\&=&-\frac{4}{3}\mathcal{M}^{a,(1);{\texttt{\tiny HBv}}}_{\mu;1(e)}\ (\texttt{provided}\ A^{(1)}=a^{(1)}=1),\\ \mathcal{M}^{a,(1);{\texttt{\tiny HBv}}}_{\mu;(f)}&=&\mathcal{M}^{a,(1);{\texttt{\tiny HBv}}}_{\mu;1(g)}=0;\\\mathcal{M}^{\texttt{\tiny HBv}}_{NN;2(b1)}&=&\langle\mathbf{\tau}_1\cdot \mathbf{\tau}_2\rangle\frac{g^2_Am^2_\pi}{128\pi^2f^4_\pi}\left\{\underline{\frac{5}{8}\rho\left(\Gamma_\epsilon+\ell_N+\frac{11}{10}\right)+\frac{5}{2}\left(\Gamma_\epsilon+\ell_N +\frac{11}{5}\right)}\overbrace{-3\left(\Gamma_\epsilon+\ell_\pi\right)}+\mathcal{R}^{\texttt{\tiny HBv}}_{2b}\right\}=\mathcal{M}^{\texttt{\tiny HBv}}_{NN;2(b2)},\\ \mathcal{M}^{\texttt{\tiny HBv}}_{NN;2(c)}&=&-\left(3-2\langle\mathbf{\tau}_1\cdot\mathbf{\tau}_2\rangle\right)\frac{g^4_Am^2_\pi}{128\pi^2f^4_\pi}\left\{\underline{\frac{45}{64}\rho \left(\Gamma_\epsilon+\ell_N+\frac{881}{270}\right)-5\frac{53}{64}\left(\Gamma_\epsilon+\ell_N+\frac{3745}{2238}\right)}\right.\nonumber\\&&\left.\overbrace{+\frac{15}{4}\left( \Gamma_\epsilon+\ell_\pi\right)}\underbrace{-\frac{12\rho}{\sigma}\arctan\sigma}+\mathcal{R}^{\texttt{\tiny HBv}}_{2c}\right\},\\\mathcal{M}^{\texttt{\tiny HBv}}_{NN;2(d)}&=&-\left( 3+2\langle\mathbf{\tau}_1\cdot\mathbf{\tau}_2\rangle\right)\frac{g^4_Am^2_\pi}{128\pi^2f^4_\pi}\left\{\underline{6\frac{33}{64}\rho\left(\Gamma_\epsilon+\ell_N+\frac{587}{834}\right) +6\frac{53}{64} \left(\Gamma_\epsilon+\ell_N+\frac{1399}{874}\right)}\right.\nonumber\\&&\left.\overbrace{-\frac{15}{4}\left(\Gamma_\epsilon+\ell_\pi\right)}+\mathcal{R}^{\texttt{\tiny HBv}}_{2d}\right\}.\eea

To make things more transparent, we have underlined all the local non-chiral terms, above-braced all the local chiral terms, and under-braced the 'anomalous' term. The residual terms are just those denoted by $\mathcal{R}$'s that vanish as $\rho\rightarrow\infty$, close examination of these terms showed that they are suppressed by $\rho^{-1/2}$. The above decomposition and classification are also summarized in Table 1 for covariant and mixed calculations.
\begin{table}[!hbp]\label{covcalc}\caption{Various components in covariant/mixed calculation}\begin{center}\begin{tabular}{ccccccc}\hline\hline Diagram&local, non-chiral&local, chiral&anomalous&suppressed\\\hline Fig.1(a,b,e)&yes/yes&yes/yes&no/no&yes/yes\\Fig.1(f,g)&yes/no&no/no&no/no&yes/no\\Fig.2(b,d)&yes/yes&yes/yes&no/no&yes/yes\\ Fig.2(c)&yes/yes&yes/yes&yes/yes&yes/yes\\\hline\hline\end{tabular}\end{center}\end{table}

Evidently, the large or non-chiral local terms and the suppressed terms $\mathcal{R}$'s in each diagram are variant with different versions of vertices, the rest are 'stable'. The anomalous term in the box diagram For $NN$ scattering (Fig. 2(c)) becomes IR singular using HB propagator for baryons\cite{Weinberg}. If the nonrelativistic propagator is used, this term becomes $\sim-3\pi$ in dimensional scheme and contaminated by linear divergence in cut-off scheme\cite{MPLA29}. It becomes definite only when covariant baryonic propagator is employed.
\subsection{Chiral 'stability' from covariant propagators}Removing all the underlined terms that are local and non-chiral, we arrive at the following 'stable' or 'preserved' components plus suppressed ones for each diagram:\bea\tilde{\mathcal{M}}^{a,(1)}_{\mu;1(a)}&=&-A^{(1)}\bar{u}\tau^a\gamma_\mu{u}\frac{g^2_Am^2_\pi}{4(4\pi f_\pi)^2}\left\{-3\left(\Gamma_\epsilon +\ell_\pi\right)+o\left(\rho^{-1/2}\right)\right\}=\frac{1}{3}\tilde{\mathcal{M}}^{a,(1)}_{\mu;1(e)},\\\tilde{\mathcal{M}}^{a,(1)}_{\mu;1(b)}&=&a^{(1)}\bar{u}\tau^a\gamma_\mu{u} \frac{g^2_Am^2_\pi}{(4\pi f_\pi)^2}\left\{-3\left(\Gamma_\epsilon+\ell_\pi\right)+o\left(\rho^{-1/2}\right)\right\},\\\tilde{\mathcal{M}}^{a,(1)}_{\mu;1(f)}&=&-A^{(1)}\bar{u}\tau^a \gamma_\mu{u}\frac{g^2_Am^2_\pi}{(4\pi f_\pi)^2}\left\{0+o\left(\rho^{-1/2}\right)\right\}=\mathcal{M}^{a,(1)}_{\mu;1(g)},\\\tilde{\mathcal{M}}_{NN;2(b1)}&=&\langle\mathbf{\tau}_1\cdot \mathbf{\tau}_2\rangle\frac{g^2_Am^2_\pi}{128\pi^2f^4_\pi}\left\{-3\left(\Gamma_\epsilon+\ell_\pi\right)+o\left(\rho^{-1/2}\right)\right\}=\mathcal{M}_{NN;2(b2)},\\ \tilde{\mathcal{M}}_{NN;2(c)}&=&-\left(3-2\langle\mathbf{\tau}_1\cdot\mathbf{\tau}_2\rangle\right)\frac{g^4_Am^2_\pi}{128\pi^2f^4_\pi}\left\{\frac{15}{4}\left(\Gamma_\epsilon +\ell_\pi\right)-\frac{12\rho}{\sigma}\arctan\sigma+o\left(\rho^{-1/2}\right)\right\},\\\tilde{\mathcal{M}}_{NN;2(d)}&=&-\left(3+2\langle\mathbf{\tau}_1\cdot\mathbf{\tau}_2\rangle \right)\frac{g^4_Am^2_\pi}{128\pi^2f^4_\pi}\left\{-\frac{15}{4}\left(\Gamma_\epsilon+\ell_\pi\right)+o\left(\rho^{-1/2}\right)\right\}.\eea

Thus, if we apply the foregoing prescription suggested on top of Eq.(\ref{mix-subtract}) (subtraction of the local non-chiral components), vector current is conserved (up to suppressed terms), and chiral power counting is ensured for the retained terms except those due to IR enhancement or threshold effects that must be kept anyway. Now it is time to recapitulate our line of arguments leading to our proposal: First, as mentioned in the introduction, there is a need to work with covariant formulation in order to avoid possible pathologies of HB or NR formulation. Second, the only source of chiral power counting violation is the mass scale of baryon/nucleon or heavy hadron, which, in addition to chiral components, results in LOCAL non-chiral pieces that are removable and possible nonlocal components of IR enhancement or threshold effects that must be kept. Third, our proposal naturally follows: isolate the local non-chiral pieces in each loop in covariant formulation and subtract them together with the supposed chiral divergences, NO need for any additional manipulations that either complicate the calculation or bring in extra pieces (like spurious power divergences in NR treatment, c.f. the end of Sec.3.2. or \cite{MPLA29}). Such a prescription could find support from the idea of 'decoupling'\cite{dec,dec2,dec3}, here it is employed to protect chiral power counting\cite{MPLA29}. It smoothly extends the standard subtraction algorithm that prevails in particle physics (say, $MS$ or $\overline{MS}$) and automatically recovers the standard one for Goldstone sector. In this sense, we may call it extended $MS$ ($\overline{MS}$), or E$MS$ (E$\overline{MS}$) for short. Specifically, according to standard algorithm for composite operator renormalization\cite{Collins} the following counterterms in for example E$\overline{MS}$ suffice to remove all the divergences and local non-chiral terms for the corresponding diagrams listed in Figure 1,\bea\delta{O}_{\mu;Cov}^{a,(1)}(N)&=&\left[\delta{Z} ^{\texttt{\tiny E}}_{O}-\delta{Z}^{\texttt{\tiny E}}_\Psi\right]{O}_{\mu;Cov}^{a,(1)}(N):\ \delta{Z}^{\texttt{\tiny E}}_\Psi=-\delta{Z}^{\texttt{\tiny E}}_{O;e},\ \delta{Z} ^{\texttt{\tiny E}}_{O}=\delta{Z}^{\texttt{\tiny E}}_{O;a}+\delta{Z}^{\texttt{\tiny E}}_{O;b}+\delta{Z}^{\texttt{\tiny E}}_{O;f+g},\\\delta{O}_{\mu;HB}^{a,(1)}(N)&=&\left[\delta{Z} ^{\texttt{\tiny E;HBv}}_{O}-\delta{Z}^{\texttt{\tiny E;HBv}}_\Psi\right]{O}_{\mu;HB}^{a,(1)}(N):\ \delta{Z}^{\texttt{\tiny E;HBv}}_\Psi=-\delta{Z}^{\texttt{\tiny E;HBv}}_{O;e},\ \delta{Z}^{\texttt{\tiny E;HBv}}_{O}=\delta{Z}^{\texttt{\tiny E;HBv}}_{O;a}+\delta{Z}^{\texttt{\tiny E;HBv}}_{O;b},\eea where the counterterms with subscripts 'a, b, e, f+g' could be readily read off from Eqs.(15-17,21-23). To illustrate, we have from Eq.(15,21) (Figure 1(a)),\bea&&\delta{Z}^{\texttt{\tiny E}}_{O;a}=\frac{g^2_Am^2_\pi}{4(4\pi f_\pi)^2} \left\{2\left(\Gamma_\epsilon+\ell_N+\frac{3}{2}\right)-3\Gamma_\epsilon\right\},\\&&\delta{Z}^{\texttt{\tiny E;HBv}}_{O;a}=\frac{g^2_Am^2_\pi} {4(4\pi f_\pi)^2}\left\{\frac{5}{2} \rho\left(\Gamma_\epsilon+\ell_N+\frac{3}{5}\right)+\frac{15}{4}\left(\Gamma_\epsilon+\ell_N+\frac{5}{3}\right)-3\Gamma_\epsilon\right\}.\eea Note that, $\delta{Z}^{\cdots}_\Psi$ is the counter term for nucleon field ($\Psi_N$) renormalization that contributes via Figure 1(e). Adding up the counterterms we have $\left[\delta{Z}^{\texttt{\tiny E}}_{O}-\delta{Z} ^{\texttt{\tiny E}}_\Psi\right]=0$ in accordance with Eq.(11). But with HB vertices, we have $$\left[\delta{Z}^{\texttt{\tiny E;HBv}}_{O}-\delta{Z}^{\texttt{\tiny E;HBv}}_\Psi\right] =\frac{5g^2_Am^2_\pi}{16(4\pi f_\pi)^2}\left\{\frac{3}{2}\rho\left(\Gamma_\epsilon+\ell_N+\frac{13}{30}\right)+\left(\Gamma_\epsilon+\ell_N+\frac{3}{5}\right)\right\}\neq0,$$ which is just the counter term needed in Eq.(12) to arrive at Eq.(13). For the $NN$ scattering diagrams in Figure 2, the counter terms are supplied by the contact $NN$ interactions that are first introduced by Weinberg in 1990\cite{Weinberg}. For those amplitudes listed in Eqs.(18-20,24-26), the counterterms generically reads\bea\delta\mathcal{L}_{NN}=-\delta{C} ^{\texttt{\tiny E}}_s\left(\bar{\Psi}_N\Psi_N\right)^2-\delta{C}^{\texttt{\tiny E}}_\tau \left(\bar{\Psi}_N{\bf\tau}\Psi_N\right)^2:\ \delta{C}^{\texttt{\tiny E}}_s=\delta{C} ^{\texttt{\tiny E}}_{s;c}+\delta{C}^{\texttt{\tiny E}}_{s;d},\ \delta{C}^{\texttt{\tiny E}}_\tau=\delta{C}^{\texttt{\tiny E}}_{\tau;b}+\delta{C}^{\texttt{\tiny E}}_{\tau;c}+\delta{C}^ {\texttt{\tiny E}}_{\tau;d}.\eea Similarly from Eq.(19,25) we have\bea\delta{C}^{\texttt{\tiny E}}_{s;c}&=&3\frac{g^4_Am^2_\pi}{128\pi^2f^4_\pi}\left\{\rho\left(\Gamma_\epsilon +\ell_N+3\right)-4\left(\Gamma_\epsilon+\ell_N+\frac{13}{16}\right)+\frac{15}{4}\Gamma_\epsilon\right\}=-\frac{3}{2}\delta{C}^{\texttt{\tiny E}}_{\tau;c},\\\delta{C}^{\texttt{\tiny E;HBv}}_{s;c}&=&3\frac{g^4_Am^2_\pi}{128\pi^2f^4_\pi}\left\{\frac{45}{64}\rho\left(\Gamma_\epsilon+\ell_N+\frac{881}{270}\right)-5\frac{53}{64}\left(\Gamma_\epsilon+\ell_N +\frac{3745}{2238}\right)+\frac{15}{4}\Gamma_\epsilon\right\}=-\frac{3}{2}\delta{C}^{\texttt{\tiny E;HBv}}_{\tau;c}.\eea

In spite that the diagrams considered here are all on-shell ones with zero three-momenta, our analysis and proposal also applies to the nonzero three momenta or off-shell cases. This is because chiral effective theories have been developed only for intermediate or low energy hadronic physics where typical momentum or momentum transfer involved must be chiral-sized, $Q\sim m_\pi\ll{M}_N$. Then, aside of the possible local terms associated with external momenta of baryons/nucleons that are non-chiral ($p^2\sim M^2_{N(B)}$) but again removable just like the non-chiral local terms associated with $M_N$, the other places that these external momenta could show up must be in non-local terms that must be kept intact, wether they are large (like threshold effects or IR enhancement) or small (like the suppressed terms $\mathcal{R}$'s). Actually, the only source of non-chiral scales are still masses of baryons/nulceons\cite{GSS} in the framework of chiral effective theories. Of course, there is no point to apply the forgoing analysis to the cases with non-chiral momentum transfer as the chiral effective theories simply break down in such circumstances.

Before closing this section, we would like to note that the prescription suggested here has already been employed in effect in some literature addressing hadronic physics using chiral effective field theories. For example, in Ref.\cite{Thomas}, the chiral components are extracted in the computation of the pion momentum distributions in nucleons with the rest parts simply discarded.
\section{Summary and prospectives}In summary, we have shown at one-loop level that the chiral components of the one-loop diagrams with one or two external baryonic/nucleonic lines are well 'preserved' as long as covariant propagators are employed for baryons/nucleons. These results naturally suggest us to advance an extended and unified prescription for chiral effective theories that dispenses extra operations and/or regularizations as long as covariant formulation is used.

As baryons still participate the dynamical processes in a 'mild or modest' sense, unlike in the standard decoupling theories where heavy particles are completely removed from the formulation of effective theories, the strategy suggested above may be seen as a modified or mitigated implementation of 'decoupling'. This, in our view, may be somehow underlying the soft cut-off or lattice regularization methods advocated in Refs.\cite{finiteC,finiteC2} for SU(3) chiral effective theory. As we have only considered the SU(2) effective theories here, it is natural to investigate if the extended prescription or strategy still works well in SU(3) cases. It is also interesting to test this extended prescription beyond one-loop level, which might provide some clues for the issue of covariant resummation of the 'anomalous' components.

\section*{Acknowledgement}We are grateful to the anonymous referee for his reports that are very helpful for improving the manuscript. JFY wishes to express his deep gratitude to F. Wang (Nanjing U) for his continuous supports and very kind helps. He is also grateful to L.-S. Geng (Beihang U), B.-W. Long (Si-Chuan U) and Dr. X.-L. Ren (Bonn U) for their interests in and suggestions to our works. This project is supported by the National Natural Science Foundation of China under Grant No. 11435005 and by the Ministry of Education of China.

\appendix
\section{Covariant calculation of one-loop diagrams}The explicit expressions of the one-loop diagrams listed in Fig. 1 and Fig. 2 will be given below. \textbf{The} completely covariant calculation yields the following\bea\mathcal{M}^{a,(1)}_{\mu;1(a)}&=&-A^{(1)}\bar{u}\tau^a\gamma_\mu{u}\frac{g^2_Am^2_\pi}{4(4\pi f_\pi)^2}\left\{-\Gamma_\epsilon-\ell_\pi-2\ln\rho+3 +\mathcal{R}_{1a}^{\texttt{\tiny c}}\right\}=\frac{1}{3}\mathcal{M}^{a,(1)}_{\mu;1(e)},\\\mathcal{M}^{a,(1)}_{\mu;1(b)}&=&a^{(1)}\bar{u}\tau^a\gamma_\mu{u}\frac{g^2_Am^2_\pi}{(4\pi f_\pi)^2}\left\{2\rho\left(\Gamma_\epsilon+\ell_N+1\right)+\Gamma_\epsilon+\ell_N-3\ln\rho+7+\mathcal{R}_{1b}^{\texttt{\tiny c}}\right\},\\\mathcal{M}^{a,(1)}_{\mu;1(c)}&=&A^{(1)} \bar{u}\tau^a\gamma_\mu{u}\frac{m^2_\pi}{(4\pi f_\pi)^2}\left(\Gamma_\epsilon+\ell_\pi+1\right),\ \mathcal{M}^{a,(1)}_{\mu;1(d)}=-\frac{m^2_\pi}{(4\pi f_\pi)^2}a^{(1)}\bar{u}\tau^a \gamma_\mu{u}\left(\Gamma_\epsilon+\ell_\pi+1\right),\\\mathcal{M}^{a,(1)}_{\mu;1(f)}&=&-A^{(1)}\bar{u}\tau^a\gamma_\mu{u}\frac{g^2_Am^2_\pi}{(4\pi f_\pi)^2}\left\{\rho\left( \Gamma_\epsilon+\ell_N+1\right)+\left(\Gamma_\epsilon+\ell_N+2\right)+\mathcal{R}_{1f}^{\texttt{\tiny c}}\right\}=\mathcal{M}^{a,(1)}_{\mu;1(g)};\\\mathcal{M}_{NN;2(b1)}&=&\langle \mathbf{\tau}_1\cdot\mathbf{\tau}_2\rangle\frac{g^2_Am^2_\pi}{128\pi^2f^4_\pi}\left\{2\rho\left(\Gamma_\epsilon+\ell_N+1\right)+\Gamma_\epsilon+\ell_N-3\ln\rho+7+\mathcal{R}_{2b} ^{\texttt{\tiny c}}\right\}=\mathcal{M}_{NN;2(b2)},\\\mathcal{M}_{NN;2(c)}&=&-\left(3-2\langle\mathbf{\tau}_1\cdot\mathbf{\tau}_2\rangle\right)\frac{g^4_Am^2_\pi}{128\pi^2f^4_\pi} \left\{\rho\left(\Gamma_\epsilon+\ell_N+3\right)-\frac{1}{4}\left(\Gamma_\epsilon+\ell_\pi+1\right)+4\ln\rho-3\right.\nonumber\\&&\left.-\frac{12\rho}{\sigma}\arctan\sigma +\mathcal{R}_{2c}^{\texttt{\tiny c}}\right\},\\\mathcal{M}_{NN;2(d)}&=&-\left(3+2\langle\mathbf{\tau}_1\cdot\mathbf{\tau}_2\rangle\right)\frac{g^4_Am^2_\pi}{128\pi^2f^4_\pi}\left\{ 3\rho\left(\Gamma_\epsilon+\ell_N+\frac{1}{3}\right)+\frac{1}{4}\left(\Gamma_\epsilon+\ell_\pi+1\right)-4\ln\rho+4+\mathcal{R}_{2d}^{\texttt{\tiny c}}\right\},\eea where $(3\pm2\langle\mathbf{\tau}_1\cdot\mathbf{\tau}_2\rangle)\equiv3\bar{u}_1u_1\bar{u}_2u_2\pm2\bar{u}_1\mathbf{\tau}_1u_1\cdot\bar{u}_2\mathbf{\tau}_2u_2$. The detailed expressions of the $\mathcal{R}_{\cdots}^{\cdots}$ terms here and below vanish in the limit $\rho\rightarrow\infty$ and will be given in Appendix C.
\section{Mixed calculation of one-loop diagrams}The calculation with HB vertices and composite operators yields the following\bea\mathcal{M}^{a,(1);{\texttt{\tiny HBv}}}_{\mu;1(a)}&=& -A^{(1)}\bar{u}\tau^av_\mu{u}\frac{g^2_Am^2_\pi}{4(4\pi f_\pi)^2}\left\{\frac{5}{2}\rho\left(\Gamma_\epsilon+\ell_N+\frac{3}{5}\right)+\frac{3}{4}\left(\Gamma_\epsilon+\ell_N +\frac{1}{3}\right)-3\ln\rho+6+\mathcal{R}^{\texttt{\tiny HBv}}_{1a}\right\},\\\mathcal{M}^{a,(1);{\texttt{\tiny HBv}}}_{\mu;1(b)}&=&a^{(1)}\bar{u}\tau^av_\mu{u}\frac{g^2_Am^2_\pi} {(4\pi f_\pi)^2}\left\{\frac{5}{8}\rho\left(\Gamma_\epsilon+\ell_N+\frac{11}{10}\right)-\frac{1}{2}\left(\Gamma_\epsilon+\ell_\pi+1\right)-\frac{5}{2}\ln\rho+6+\mathcal{R} ^{\texttt{\tiny HBv}}_{1b}\right\}\nonumber\\&=&-\frac{4}{3}\mathcal{M}^{a,(1);{\texttt{\tiny HBv}}}_{\mu;1(e)}\ (\texttt{provided}\ A^{(1)}=a^{(1)}=1),\\\mathcal{M}^{a,(1); {\texttt{\tiny HBv}}}_{\mu;1(c)}&=&A^{(1)}\bar{u}\tau^av_\mu{u}\frac{m^2_\pi}{(4\pi f_\pi)^2}\left(\Gamma_\epsilon +\ell_\pi+1\right),\ \mathcal{M}^{a,(1);{\texttt{\tiny HBv}}}_{\mu;1(d)}=-a^{(1)}\bar{u}\tau^av_\mu{u} \frac{m^2_\pi}{(4\pi f_\pi)^2}\left(\Gamma_\epsilon+\ell_\pi+1\right),\\\mathcal{M}^{a,(1);{\texttt{\tiny HBv}}}_{\mu;(f)}&=& \mathcal{M}^{a,(1);{\texttt{\tiny HBv}}}_{\mu;1(g)}=0,\\ \mathcal{M}^{\texttt{\tiny HBv}}_{NN;2(b1)}&=&\langle\mathbf{\tau}_1\cdot\mathbf{\tau}_2\rangle\frac{g^2_Am^2_\pi} {128\pi^2f^4_\pi}\left\{\frac{5}{8}\rho\left(\Gamma_\epsilon+\ell_N+\frac{11}{10}\right)-\frac{1}{2}\left(\Gamma_\epsilon+\ell_\pi+1\right)-\frac{5}{2}\ln\rho+6+\mathcal{R}^ {\texttt{\tiny HBv}}_{2b}\right\}\nonumber\\&=&\mathcal{M}^{\texttt{\tiny HBv}}_{NN;2(b2)},\\\mathcal{M}^{\texttt{\tiny HBv}}_{NN;2(c)}&=&-\left(3-2\langle\mathbf{\tau}_1 \cdot\mathbf{\tau}_2\rangle\right)\frac{g^4_Am^2_\pi}{128\pi^2f^4_\pi}\left\{\frac{45}{64}\rho\left(\Gamma_\epsilon+\ell_N+\frac{881}{270}\right)-\frac{133}{64}\left(\Gamma_\epsilon +\ell_N\right)+\frac{15}{4}\ln\rho-\frac{3745}{384}\right.\nonumber\\&&\left.-\frac{12\rho}{\sigma}\arctan\sigma+\mathcal{R}^{\texttt{\tiny HBv}}_{2c}\right\},\\\mathcal{M}^ {\texttt{\tiny HBv}}_{NN;2(d)}&=&-\left(3+2\langle\mathbf{\tau}_1\cdot\mathbf{\tau}_2\rangle\right)\frac{g^4_Am^2_\pi}{128\pi^2f^4_\pi}\left\{\frac{417}{64}\rho\left(\Gamma_\epsilon +\ell_N+\frac{587}{834}\right)+\frac{197}{64}\left(\Gamma_\epsilon+\ell_N\right)-\frac{15}{4}\ln\rho+\frac{1399}{128}\right.\nonumber\\&&\left.+\mathcal{R}^{\texttt{\tiny HBv}}_{2d}\right\}.\eea
\section{Definitions of $\mathcal{R}$'s}\bea\mathcal{R}_{1a}^{\texttt{\tiny c}}&\equiv&\frac{4-12\rho}{\rho\sigma}\arctan\sigma+2\frac{\ln\rho}{\rho},\\\mathcal{R}_{1b}^{\texttt{\tiny c}}&\equiv&\frac{6-20\rho}{\rho\sigma}\arctan\sigma+3\frac{\ln\rho}{\rho}=\mathcal{R}_{2b}^{\texttt{\tiny c}},\\\mathcal{R}_{1f}^{\texttt{\tiny c}}&\equiv&\frac{ -\sigma\arctan\sigma}{\rho}+\frac{\ln\rho}{2\rho},\\\mathcal{R}_{2c}^{\texttt{\tiny c}}&\equiv&\frac{1}{\rho}\left(\frac{14\rho-3}{\sigma}\arctan\sigma-\frac{3}{2}\ln\rho\right),\\ \mathcal{R}_{2d}^{\texttt{\tiny c}}&\equiv&\left(-15+\frac{3}{\rho}-\frac{1}{\sigma^2}\right)\frac{\arctan\sigma}{\sigma}+\frac{3\ln\rho}{2\rho}+\frac{1}{\sigma^2},\\\mathcal{R} ^{\texttt{\tiny HBv}}_{1a}&\equiv&\frac{5}{4\rho}\left(\frac{(1-4\rho)\sigma}{\rho}\arctan\sigma+\frac{6\rho-1}{2\rho}\ln\rho-1\right),\\\mathcal{R}^{\texttt{\tiny HBv}}_{1b}&\equiv& \frac{5}{16\rho}\left(\frac{ (-2+12\rho-16\rho^2)\sigma}{\rho^2}\arctan\sigma+\frac{1-8\rho+18\rho^2}{\rho^2}\ln\rho-\frac{13\rho-2}{\rho}\right)=\mathcal{R}^{\texttt{\tiny HBv}}_{2b}, \\\mathcal{R}^{\texttt{\tiny HBv}}_{2c}&\equiv&\left(30-\frac{19}{\rho}+\frac{31}{8\rho^2}-\frac{1}{64\rho^3}-\frac{3}{64\rho^4}\right)\frac{\arctan\sigma}{\sigma} +\left(-\frac{393}{64}+\frac{57}{32\rho}-\frac{7}{128\rho^2}-\frac{3}{128\rho^3}\right)\frac{\ln\rho}{\rho}\nonumber\\&&+\frac{415}{128\rho}-\frac{23}{128\rho^2}-\frac{3}{64\rho^3},\\ \mathcal{R}^{\texttt{\tiny HBv}}_{2d}&\equiv&\left(-35+\frac{133}{4\rho}-\frac{5}{16\rho^2}-\frac{121}{64\rho^3}-\frac{23}{64\rho^4}-\frac{1}{32\rho^4\sigma^2}\right) \frac{\arctan\sigma}{\sigma}+\left(\frac{591}{64}-\frac{85}{32\rho}-\frac{155}{128\rho^2}-\frac{21}{128\rho^3}\right)\frac{\ln\rho}{\rho}\nonumber\\&&-\frac{1029}{128\rho} -\frac{309}{128\rho^2}-\frac{13}{64\rho^3}+\frac{1}{8\rho^3\sigma^2}.\eea


\begin{thebibliography}{00}
\bibitem{Isgur}N. Isgur, M.B. Wise, Phys. Lett. \textbf{B232}, 113 (1990).
\bibitem{Jenkins}E. Jenkins, A. Manohar, Phys. Lett. \textbf{B255}, 558 (1991), {\em ibid}, \textbf{B259}, 353 (1991).
\bibitem{Bernard}V. Bernard, Prog. Part. Nucl. Phys. \textbf{60}, 82 (2008) and references therein.
\bibitem{Weinberg}S. Weinberg, Phys. Lett. \textbf{B251}, 288 (1990).
\bibitem{vK}C. Ordonez, L. Ray, U. van Kolck, Phys. Rev. Lett. \textbf{72}, 1982 (1994).
\bibitem{RevB}S.R. Beane, {\em et al}, arXiv: nucl-th/0008064.
\bibitem{RevBvK}P.F. Bedaque, U. van Kolck, Ann. Rev. Nucl. Part. Sci. \textbf{52}, 339 (2002).
\bibitem{RevEpel}E. Epelbaum, Prog. Part. Nucl. Phys. \textbf{57}, 654 (2006).
\bibitem{RevRMP81}E. Epelbaum, H.-W. Hammer, Ulf-G. Mei\ss ner, Rev.Mod. Phys. \textbf{81}, 1773 (2009).
\bibitem{RevRP503}R. Machleidt, D.R. Entem, Phys. Report \textbf{503}, 1 (2011).
\bibitem{GSS}J. Gasser, M.E. Sainio, A. Svarc, Nucl. Phys. \textbf{B307}, 779 (1988).
\bibitem{GL}J. Gasser, H. Leutwyler, Ann. Phys. \textbf{158}, 142 (1984).
\bibitem{GL2}J. Gasser, H. Leutwyler, Nucl. Phys. \textbf{B250}, 465 (1985).
\bibitem{becher}T. Becher, H. Leutwyler, Eur. Phys. J. \textbf{C9}, 643 (1999).
\bibitem{PVRA}M.P. Valderrama, E.R. Arriola, Phys. Rev. \textbf{C72}, 044007 (2005), {\em ibid}, \textbf{C74}, 054001, 064004 (2006).
\bibitem{HPVRA}R. Higa, M.P. Valderrama, E.R. Arriola, Phys. Rev. \textbf{C77}, 034003 (2008).
\bibitem{EpelMeis}E. Epelbaum, Ulf-G. Mei\ss ner, arXiv: nucl-th/0609037.
\bibitem{EG}E. Epelbaum, J. Gegelia, Eur. Phys. J. \textbf{A41}, 341 (2009).
\bibitem{MLEA}R. Machleidt,{\em et al}, Phys. Rev. \textbf{C81}, 024001 (2010).
\bibitem{LongYang}B. Long, C.-J. Yang, Phys. Rev. \textbf{C84}, 057001 (2011), {\em ibid}, \textbf{C85}, 034002, \textbf{C86}, 024001 (2012).
\bibitem{ZME}Ch. Zeoli, R. Machleidt, D.R. Entem, Few Body Syst. \textbf{54}, 2191 (2013).
\bibitem{EG716}E. Epelbaum, J. Gegelia, Phys. Lett. \textbf{B716}, 338 (2012).
\bibitem{MPLA29}J.-F. Yang, Mod. Phys. Lett. \textbf{A29}, 1450043 (2014).
\bibitem{Gege-Scherer}T. Fuchs, {\em et al}, Phys. Rev. \textbf{D68}, 056005 (2003).
\bibitem{Beihang}X.L. Ren, {\em et al}, Chin.Phys. \textbf{C42}, 014103 (2018).
\bibitem{dec}T. Appelquist, J. Carazzone, Phys. Rev. \textbf{D11}, 2856 (1975).
\bibitem{dec2}E. Witten, Nucl. Phys. \textbf{B104}, 445 (1976).
\bibitem{dec3}S. Weinberg, Phys. Lett. \textbf{B91}, 51 (1980).
\bibitem{JWCJi}J.-W. Chen, X.-D. Ji, Phys. Lett. \textbf{B523}, 73, 107 (2001).
\bibitem{Arndt}D. Arndt, M.J. Savage, Nucl. Phys. \textbf{A697}, 429 (2002).
\bibitem{Collins}See, e.g., J.C. Collins, {\em Renormalization}, Cambridge University Press (1984).
\bibitem{Thomas}M. Burkhardt, {\em et al}, Phys. Rev. \textbf{D87}, 056009 (2013).
\bibitem{finiteC}J.F. Donoghue, B.R. Holstein, Phys. Lett. \textbf{B436}, 331 (1998).
\bibitem{finiteC2}J.F. Donoghue, B.R. Holstein, B. Borasoy, Phys. Rev. \textbf{D59}, 036002 (1999).
\end{thebibliography}
\end{document}